\documentclass[aps,twocolumn,prb,floatfix,superscriptaddress,showpacs,longbibliography]{revtex4-1}
\usepackage[utf8]{inputenc}
\usepackage{amssymb}
\usepackage{epsf}
\usepackage{epsfig}
\usepackage{graphicx}
\usepackage{dcolumn}
\usepackage{braket}
\usepackage{bm}
\usepackage{amsfonts}
\usepackage{amsmath}
\usepackage{amssymb}
\usepackage{color,soul}
\usepackage{wasysym}
\usepackage{mathrsfs}
\usepackage{times}
\usepackage[dvipsnames,svgnames,table]{xcolor}
\usepackage{ulem}
\usepackage{hyperref}
\usepackage{fancyhdr}
\usepackage[english]{babel}
\usepackage{hyperref}
\usepackage{float}

\hypersetup{
    unicode=true,          % non-Latin characters in Acrobat’s bookmarks
    pdftoolbar=true,        % show Acrobat’s toolbar?
    pdfmenubar=true,        % show Acrobat’s menu?
    pdffitwindow=false,     % window fit to page when opened
    pdfstartview={FitH},    % fits the width of the page to the window
    pdftitle={2017-Dal Lago et al. - Floquet isolator},    % title
    pdfauthor={LEFFT},     	% author
    colorlinks=true,       	% false: boxed links; true: colored links
    linkcolor=NavyBlue,  % color of internal links (change box color with linkbordercolor)
    citecolor=Maroon,       % color of links to bibliography
    filecolor=NavyBlue,      % color of file links
    urlcolor=NavyBlue       % color of external links
}

\newcommand{\bea}{\begin{eqnarray}}
\newcommand{\eea}{\end{eqnarray}}

\newcommand{\ci}{i}

\begin{document}

\title{One-way transport in laser-illuminated bilayer graphene:\\ A Floquet isolator}

\author{V. Dal Lago}
\email{vdallago@famaf.unc.edu.ar}
\affiliation{Instituto de F\'{\i}sica Enrique Gaviola (CONICET) and FaMAF, Universidad Nacional de C\'ordoba, Argentina}
\author{E. Su\'arez Morell}
\affiliation{Departamento de F\'{\i}sica, Universidad T\'ecnica Federico Santa Mar\'{\i}a, Valpara\'{\i}so, Chile}
\author{L. E. F. Foa Torres}% $^*$}
\email{luis.foatorres@uchile.cl}
\affiliation{Departamento de F\'{\i}sica, Facultad de Ciencias F\'{\i}sicas y Matem\'aticas, Universidad de Chile, Santiago, Chile}

\begin{abstract}
We explore the Floquet band-structure and electronic transport in laser-illuminated bilayer graphene. By using a bias voltage perpendicular to the graphene bilayer we show how to get one-way charge and valley transport among two unbiased leads. In contrast to quantum pumping, our proposal uses a different mechanism based on generating a non-reciprocal bandstructure with a built-in directionality. The Floquet states at one edge of a graphene layer become hybridized with the continuum on the other layer, and so the resulting bandstructure allows for one-way transport as in an \textit{isolator}. Our proof-of-concept may serve as a building block for devices exploiting one-way states.
\end{abstract}
\pacs{ 73.20.At; 03.65.Vf; 72.80.Vp}
%73.20.At Surface states, band structure, electron density of state
%03.65.Vf: Phases: geometric; dynamic or topological
%72.80.Vp Electronic transport in graphene

\date{\today}
\maketitle

\section{Introduction}

\begin{figure}[tbp]
\centering
\includegraphics[width=0.9\columnwidth, angle=0]{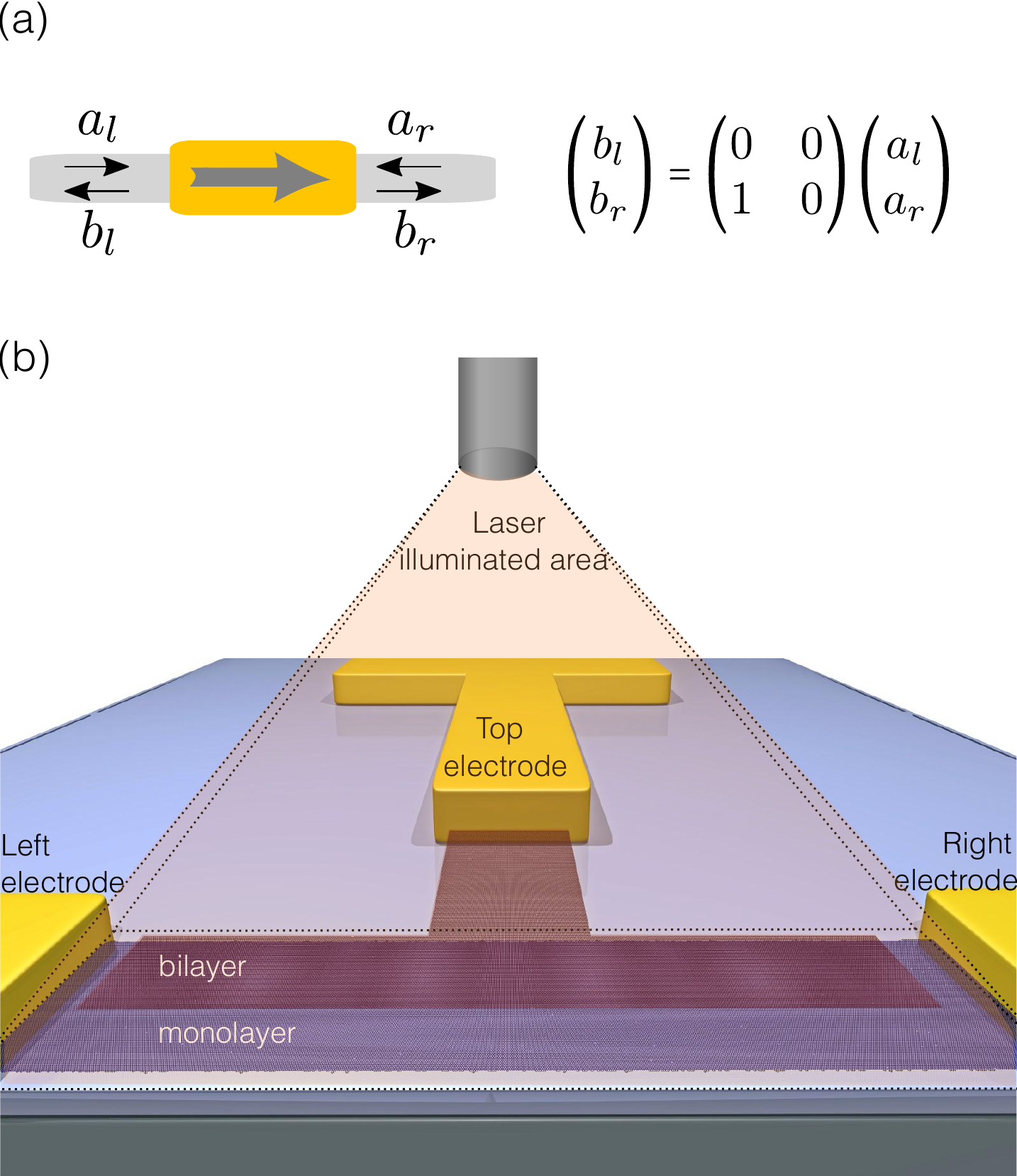}
\caption{(color online) (a) Scheme of the simplest ideal isolator (as discussed in Ref.~\onlinecite{Jalas2013}) where single moded wires are connected to a sample. The scattering matrix is skewed and transmission occurs only from left to right. 
(b) Scheme of the setup considered in this work to obtain an isolator effect: A laser illuminated graphene sample connected to left and right electrodes, an inhomogenous region with monolayer and bilayer graphene, and a third electrode on top. The bilayer area is biased perpendicularly to the graphene plane.}
\label{fig1}
\end{figure}

The advent of graphene~\cite{Novoselov2004a,Novoselov2005a,Zhang2005,CastroNeto2009} as well as the new family of two-dimensional (2D) materials and their heterostructures~\cite{Geim2013} has provided us with an outstanding playground for testing quantum transport concepts and ideas, from devices exploiting quantum interference~\cite{Rickhaus2013,Grushina2013,Gehring2016} to schemes harnessing the valley degree of freedom~\cite{Schaibley2016}. In spite of the rapid progress, controlling or steering the flow of charge, spin and valley currents across a device or material has remained as a main challenge. A promising control path is achieving \textit{one-way} transport, a situation where the current (of charge, valley or spin) can flow among two electrodes in one direction only. An example is the recent realization of non-reciprocal supercurrent flow in a carbon nanotube~\cite{Qin2017}. An even more challenging path is seeking for one-way transport in an isolator configuration. The isolator concept is borrowed from photonics~\cite{Jalas2013} and is used here for a setup where transport is possible only from lead $L$ to lead $R$ but not in the opposite direction while the reflection at lead $L$ vanishes (Fig. \ref{fig1} (a)). As the scattering matrix associated to the two terminals in an isolator is non-unitary, achieving an isolator for electrons might seem impossible at first sight, but, as will become clear later, this is not the case.

Here we show a proof-of-concept for an electronic isolator obtained by shining a laser on bilayer graphene. A previous study~\cite{FoaTorres2016} predicted a scheme for realizing this effect for charge (spin) transport on a biased bilayer graphene by including a Haldane (spin-orbit) term. In spite of the difficulty to realize a Haldane term~\cite{Haldane1988} in a condensed matter experiment, our results show that laser illumination can take a similar role and use it to produce a targetted non-reciprocity. 
This proposal is, to the best of our knowledge, the first one for an electronic/valley isolator using a time-dependent field. By offering a proof of concept, our work thus paves the way for new optoelectronic devices.

In the following we deepen on the isolator concept and provide further motivation to this work before turning to our model and the results.

\textit{The isolator concept in optics.} In optics, a device allowing light to pass in one direction but blocking it in the opposite one is called an \textit{isolator}~\cite{Jalas2013}. It is of great use in photonics where reflections that may, for example, reveal information to an observer intercepting the signal, are undesired. Moreover, in an isolator, other effects such as spurious interferences and light rerouting can be lessened if not elliminated. Theoretically, it requires at least two single-mode terminals connected to the device in such a way that transmission can occur from one terminal to the other, but vanishes in the reciprocal direction. As noted in Ref.~\onlinecite{Jalas2013}, the isolator has either to perfectly block the transmission in one direction or divert it to a third terminal. Thus the scattering matrix that represents an isolator must be asymmetric. For a two terminal system having a single channel on each terminal an isolator would be as shown in Fig.~\ref{fig1}(a).

\textit{Opportunities and challenges for the isolator concept in electronics.}
An isolator-like effect is also desirable in other contexts such as for example in electronics where directional transport is useful for logical applications, yet, to the best of our knowledge, it is still missing from the toolkit of quantum devices. The first challenge is immediately apparent: in typical electronic devices the scattering matrix is warranted to be unitary and as such a skewed matrix like the one mentioned earlier is not possible. %, at least for a system with two electrodes.
 Notwithstanding, this is not an obstacle if we consider a system with a third electrode intended solely to reroute or divert the charges propagating in one direction. The effective scattering matrix for the two terminal system may indeed be skewed in such case without violating unitarity. But even when these issues can be circumvented by adding a third physical terminal, it remains the challenge on how to get a perfectly non-recyprocal system, one where transmission from $L$ to $R$ is one and zero from $R$ to $L$.

If we think of systems hosting perfectly transmitting and robust states, the first that comes to mind is a sample hosting chiral states like in the quantum Hall regime. But this also occurs in others non-equilibrium situations~\cite{Oka2009,Lindner2011,Cayssol2013} such as in the case of graphene irradiated with a laser, where Floquet chiral edge states emerge in the spectrum~\cite{Oka2009,Kitagawa2011,Perez-Piskunow2014,Kundu2014,Dehghani2014,Gomez-Leon2014,Lopez2015,Du2017}. Since the bandstructure in such systems is recyprocal, to get an isolator-like behavior one would need to supress transport through the states propagating along one of the edges. However this is prevented by the bulk-boundary correspondence, thereby requiring a scheme to circumvent it.

\textit{This work.} Here we offer a proof-of-concept where an isolator effect is obtained in a laser-illuminated graphene bilayer. The proposed system has two `active' electrodes connected to one of the layers (which form the isolator) and a third electrode connected to the other layer which is used for diverting unwanted reflections (see scheme in Fig.~\ref{fig1}(b)). As it will become clear later on, since driving has a crucial role as it promotes the non-reciprocity for the electronic/valley transport, we call this device a \textit{Floquet isolator}.

We should also comment on the difference with the large body of previous works aimed at obtaining directional charge transport using time-dependent fields. This includes, notably, the phenomenon called quantum pumping, the use of time-dependent potentials to steer transport and obtain a dc current at zero bias or even against an external bias voltage. In the open regime~\cite{Brouwer1998,Altshuler1999,Kaestner2015} quantum pumping usually relies on quantum interference and has been extensively studied for both the adiabatic~\cite{Brouwer1998,Prada2009,Zhu2009,Ingaramo2013,Khani2016} and non-adiabatic cases~\cite{Kaestner2015,Arrachea2005,FoaTorres2005,FoaTorres2011,Croy2012}. The proposal in this paper is different from those in different aspects. First, in our case the directionality is built-in the electronic structure of the system, and second, the scattering matrix for a quantum pump does not generally have a perfect directionality as in an isolator. 

Finally, there is also one interesting previous study~\cite{Lira2012} where optical non-reciprocity is produced through electrical driving, this is the converse of our proposal where non-reciprocity in charge or valley transport is produced through laser illumination. Another recent work aiming at tailoring non-reciprocity but in interacting systems is Ref.~\onlinecite{Morimoto2017}. In our case we use laser illumination and take non-reciprocity to the extreme of producing an almost perfect isolator.

%------------------------------------------------------------------------------------
%------------------------------------------------------------------------------------

\section{Hamiltonian model and Floquet solution scheme}

%\begin{widetext}

\begin{figure*}[tbp]
\centering
\includegraphics[width=1.0\textwidth, angle=0]{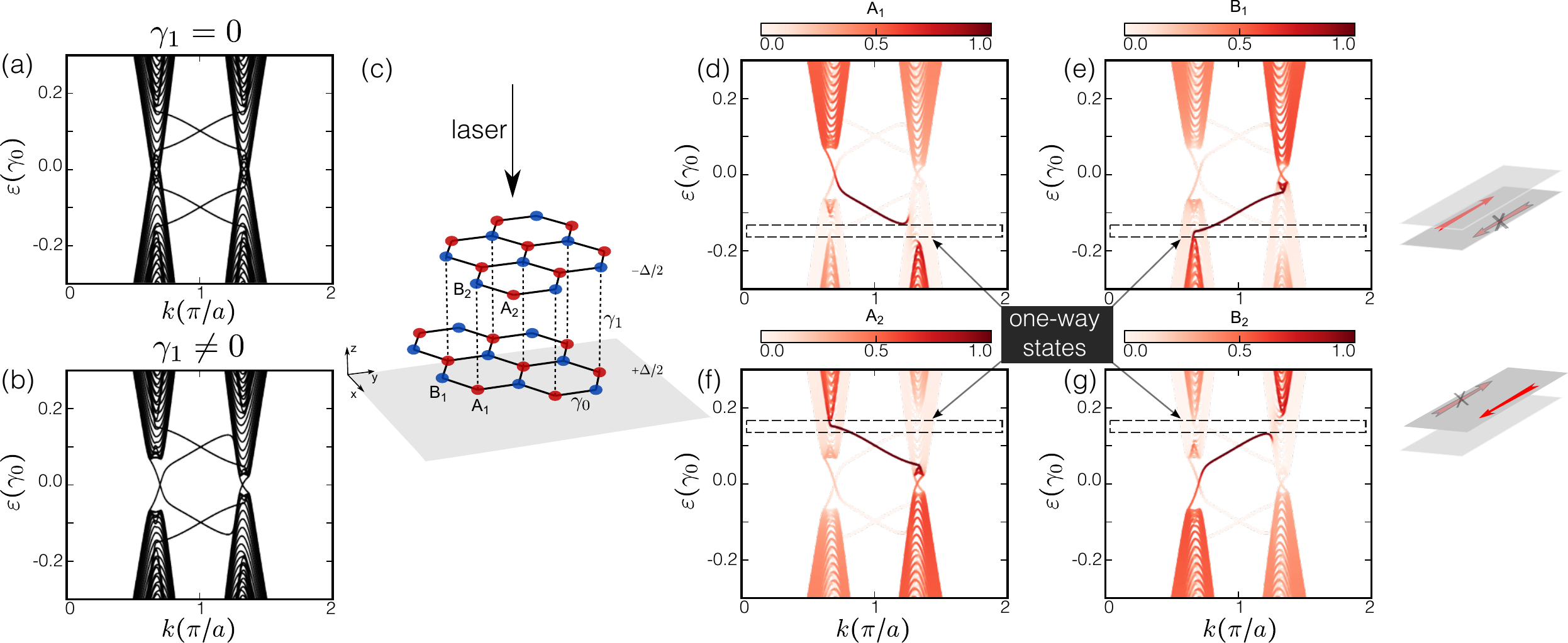}
\caption{(color online) (a) and (b) The full dispersion for the electronic states of a bilayer graphene ribbon irradiated with a laser, with the interlayer coupling off and on respectively. (c) Scheme of a graphene bilayer with Bernal stacking and a bias voltage applied perpendicularly. Panels (d) - (g) correspond to the same dispersion with a color scale encoding the weight of the states on sites A$_1$, B$_1$, A$_2$ and B$_2$ for the Floquet replica $n=0$. The parameters are chosen with the aim of illustrating the proposed mechanism: $\Delta = 0.2$, $W=103a$, $\hbar\Omega=3.5$ and $A_0=0.5 \Phi_0/(2\pi a)$.}
\label{fig2}
\end{figure*}

To motivate our discussion we consider a simple Hamiltonian for the electronic excitations in bilayer graphene:

\begin{equation}
\label{Hamiltonian}
{\cal H}=\sum_{i}E_{i}^{{}}\,c_{i}^{\dagger}c_{i}^{{}}-\sum_{\left\langle i,j\right\rangle} \gamma_{i,j} c_{i} ^{\dagger}c_{j}^{{}}+{\cal H}_{\perp},
\end{equation}

\noindent where $c_{i}^{\dagger}$ and $c_{i}^{{}}$ are the electronic creation and annihilation operators at the $\pi$-orbital on site $i$ (which can be A-type or B-type, $A_1$ and $B_1$ for the lower layer and $A_2$ and $B_2$ for the upper one). The second summation runs over nearest-neighbors, and the associated hopping matrix elements $\gamma_{i,j}$ are taken all equal to $\gamma_{0}= 2.7$ eV ~\cite{CastroNeto2009}, which is considered as the unit of energy hereafter. To model a bias voltage applied perpendicularly to the graphene bilayer we include a shift of the site energies on the lower ($E_{i}=E_{0}-\Delta/2$) and upper ($E_{i}=E_{0}+\Delta/2$) layers. Without loss of generality we take $E_{0}=0$. We consider a bilayer graphene with Bernal stacking (see Fig. \ref{fig2}(c)). In our model we consider the hopping matrix elements between $A_1$ and $B_2$ sites, $\gamma_{1}$, which is included in the ${\cal H}_{\perp}$ of the Hamiltonian.

It has been shown that laser-illuminated graphene monolayers~\cite{Oka2009,Gu2011,Perez-Piskunow2014,Du2017} and bilayers~\cite{SuarezMorell2012,Qu2017,Iorsh2017} may host chiral edge states if the laser parameters are appropriately chosen~\cite{Perez-Piskunow2015}. These Floquet chiral edge states will be the `substrate' on which we will base our proposal.

Here, we consider a sample irradiated with a circularly polarized laser perpendicular to the bilayer. This is incorporated in the Hamiltonian through the Peierls substitution as a time dependent phase in the nearest-neighbors matrix elements\cite{Calvo2013,Calvo2012}:

\begin{equation}
\label{gammat}
{\gamma_{ij}(t)} = \gamma_{0} \exp \left[ {\rm i} \cdot \frac{2\pi}{\Phi_{0}} \int_{\mathbf{r}_j}^{\mathbf{r}_i} \mathbf{A}(t) \ d\mathbf{r} \right].
\end{equation}

\noindent where $\mathbf{A}(t)$ is the vector potential of the radiation and ${\Phi_{0}}$ the magnetic flux quantum. As it can be noticed, $\gamma(t)$ depends on the position of the two nearest-neighbors sites under consideration. Then, since the radiation is perpendicular to the graphene plane, the alteration of $\gamma_{1}$ can be despised. Particularly, the vector potential associated with a circularly polarized monochromatic plane wave in the $z$-direction (perpendicular to the graphene sheet) is $\mathbf{A}(t) = A_{0} \left( \sin (\Omega t) \ \hat{x} +  \cos (\Omega t) \  \hat{y}  \right)$, where $A_{0}$ is related to the driving amplitude and $\Omega$ corresponds to the radiation frequency.

The spectral and transport properties of this system can be calculated from the time independent Floquet Hamiltonian (${\cal{H}}_F = {\cal H}- {\rm i} \hbar\partial_t$). This theory is a suitable approach due to the time-periodicity ($2\pi/\Omega$ in our case) of the Hamiltonian. For a given ${\cal{H}}_F$ the eigenvalue problem, analogous to that of the time-independent Schr\"odinger equation, can be solved in the Floquet space. This space is defined as the direct product ($\mathscr{R}\otimes \mathscr{T}$) between the usual Hilbert space ($\mathscr{R}$) and the one of the time periodic functions ($\mathscr{T}$) \cite{Sambe1973}. Particularly, there is a complete set of Floquet solutions of the form $\psi_{\alpha}(\bm{r},t)=\exp(-\ci\varepsilon_{\alpha}t/\hbar) \phi_{\alpha}(\bm{r},t)$, where ${\varepsilon_{\alpha}}$ are the so-called quasienergies and $\phi_{\alpha}(\bm{r},t+T)=\phi_{\alpha}(\bm{r},t)$ the associated Floquet states. The $\mathscr{T}$ space is spanned by the $\exp(i n\Omega t)$ functions, where the index $n$ may be interpreted as the number of `photon' excitations  and defines the $n^{th}$ Floquet replica subspace \cite{Shirley1965}.

The calculations presented in the following sections were carried out with home-made codes built on the Kwant~\cite{Groth2014} module.

%------------------------------------------------------------------------------------
%------------------------------------------------------------------------------------

\section{Generating a non-recyprocal Floquet bandstructure with unbalanced chiral edge states}

To design an isolator device we take advantage of the chiral edge states found in irradiated graphene~\cite{Perez-Piskunow2014,Usaj2014,Perez-Piskunow2015} and look for a way of annihilating one of those. Specifically, we aim to generate a non-reciprocal bandstructure with unbalanced chiral states where directional transport can be achieved.

In references \onlinecite{Usaj2014,Perez-Piskunow2015}, the authors have shown that when illuminating a graphene monolayer with a laser, a gap at zero energy opens in the dispersion relation (besides others at energies which are integer multiples of the laser frequency and half the laser frequency), and chiral states propagating along the edges emerge. 

If we take the interlayer coupling of the bilayer graphene sample to be zero, then applying a perpendicular bias leads to a bandstructure that is equivalent to that of two laser-illuminated monolayers but shifted in energy (as shown in Fig. \ref{fig2}(a)). Notably, on each monolayer one observes the opening of a gap around the Dirac point and edge states bridging them. When the interlayer coupling is turned on, one gets the results shown in figure \ref{fig2}(b). Besides a valley asymmetry due to the breaking of inversion symmetry, we see that the edge states bridging the laser-induced gap seem to remain almost intact.

A more compelling feature is hidden in this dispersion relation but can be revealed when coloring it according to the weight on the different sites of the bilayer. Figures \ref{fig2} (d)-(g) present the spectrum with a color scale encoding the weight in the lower ((d) for site $A_1$ and (e) for $B_1$) and upper (sites $A_2$ in (f) and $B_2$ (g)) layers. It can be noticed that the emerging edge states in the zero energy gap have a strong sublattice polarization (greater than 95$\%$). The origin of this polarization is the radiation itself (which is responsible of the appearance of the chiral edge states) and it is also favored by the presence of the perpendicular bias.

However, the most significative characteristic exposed in figures \ref{fig2}(d) and (g) is the existence of a region of energy where the (weight of the) chiral edge state vanishes. This disappearance is originated in the hybridization between the edge states in one layer and the continuum on the other one in the graphene sample. Specifically, due to the chosen Bernal stacking, sites $A_1$ interact with the $B_2$ through $\gamma_1$ (shown in figure \ref{fig2}(c)). Even though the value of $\gamma_1$ is small ($\gamma_1=0.1\gamma_0$),  due to the high density of states in the metallic layer (at the energy of interest), it is enough to `switch off' the chiral edge state. On the other hand, the states polarized in $A_2$ or $B_1$ are more protected to the hybridization (due to the lack of a direct interaction) and so more robust. As stated before, this annihilation of only one of the chiral edge states was our focus in order to build an isolator.

The radiation and the bias have also direct influence on where this effect is observed in the dispersion relation. In particular by switching the polarization of one of them the valley in which the disappearance occurs can be swapped.

%------------------------------------------------------------------------------------
%------------------------------------------------------------------------------------

\section{Transport properties, one-way charge transport and isolator effect}

\begin{figure}[tbp]
\centering
\includegraphics[width=0.9\columnwidth, angle=0]{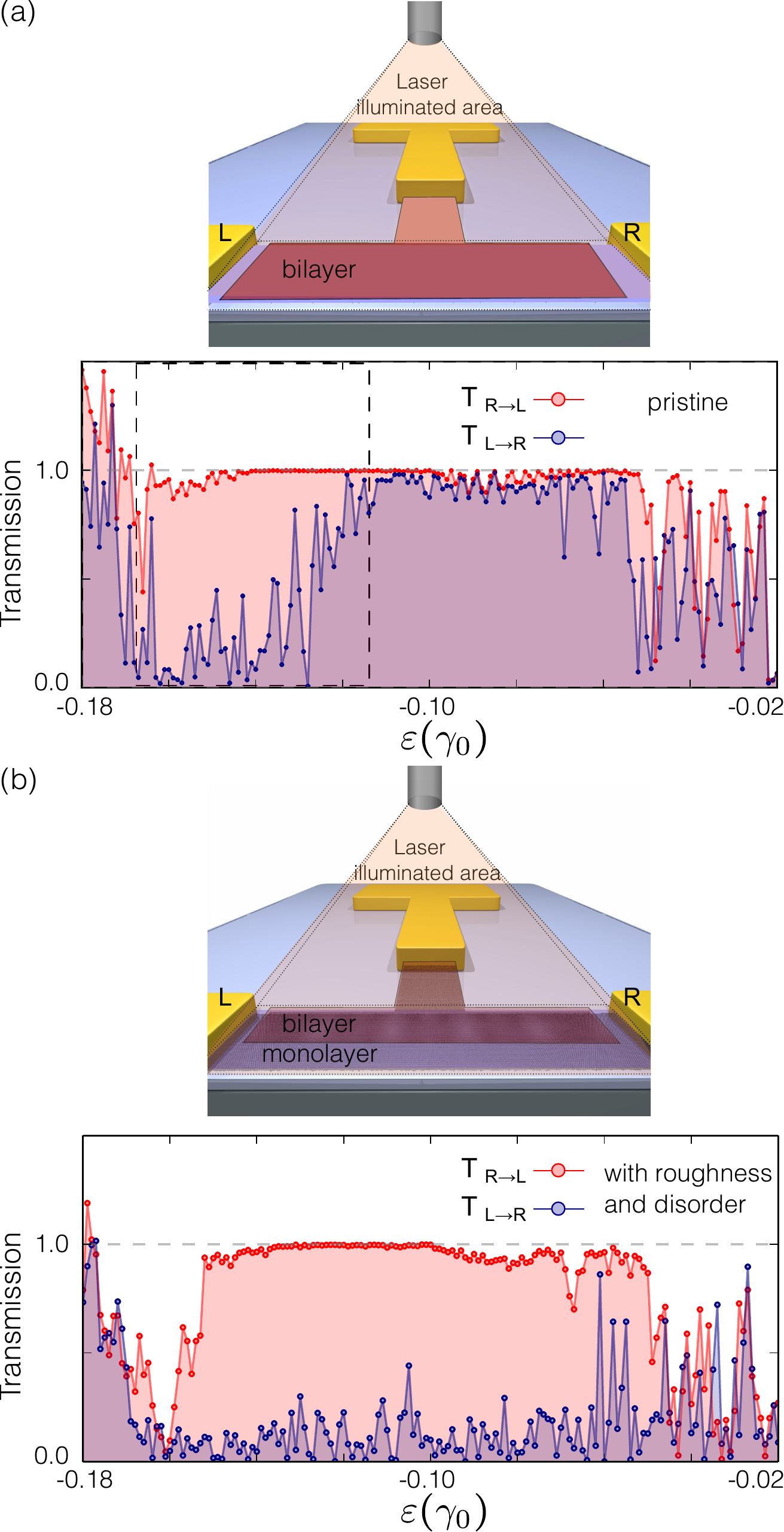}
\caption{(color online) (a) (top) Scheme of the three terminal setup where a bilayer sample is connected to three monolayer graphene leads, two of them on the lower layer and a third one to the upper layer. (bottom) Transmission probabilities between the left (L) and right (R) leads as a function of the electronic energy. One can see a strong directionality in the energy region marked with a dashed box in Fig. 2(d) and (g). (b) Same as (a) for the case when the upper layer covers only half of the lower one, with rough edges and $1\permil$ vacancies. In this case directional transport is achieved throughout the full gap as explained in the text.}
\label{fig3}
\end{figure}

Illuminating the bilayer graphene sample with a circularly polarized laser breaks the time reversal symmetry. The same result is attained on the inversion symmetry due to the external bias. Hence, the right-left symmetry is broken in our system. However, this cannot be exploited when connecting only two leads to the sample. The unitary of the $2 \times 2$ scattering matrix implies that the transmission from the left lead to the right one is equal to the transmission in the opposite direction. To elude this obstacle in the search of a \textit{Floquet isolator}, a third monolayer graphene lead is incorporated by attaching it to the upper layer of the sample (see schemes on Fig. \ref{fig3}(a) and (b)). The unitary is now compulsory for the new $3 \times 3$ scattering matrix, whereas the effective $2 \times 2$ scattering matrix can be asymmetric allowing us to profit from the non-recyprocity of the dispersion relation.

In figure \ref{fig3}(a) we show the left to right and right to left transmission probablities for the irradiated bilayer graphene sample. The chosen range of energy corresponds to the one at which one of the chiral edge states vanishes in the dispersion relation (Fig. \ref{fig2}(d) and (g)). We assumed that the thermalization process occurs in the leads as is usual in Floquet scattering theory.~\cite{Gu2011,Foa2014} The figure shows how the non-reciprocity of the band structure is reflected as a directional asymmetry in the transmissions.

Comparing our results to the ones of an ideal isolator (fig. \ref{fig1}), we can notice that they differ in the $T_{R \rightarrow L}$ which is not perfectly zero though $T_{L \rightarrow R}$ is almost one. Two main factors contribute to this small $T_{L \rightarrow R}$: a) the unavoidable interface between illuminated and non-illuminated areas which may lead to additional scattering (and which indeed produces a reflection filtering all but one transmission channel), b) the abrupt change from the bilayer sample to the graphene monolayer leads. Nonetheless, the directionality is still evident.

Ideally, an electronic isolator should be robust to disorder and roughness. This implies that its properties (like the directional asymmetry), shall not be modified by details such as edge ending or roughness, or the presence of defects. However, to produce the isolator-like behavior we relied on the sublattice polarization of the Floquet chiral states (see Fig. \ref{fig2}(d)-(g)) which allows for the upper layer to introduce a selective `environment' to the lower one thanks to the stacking order. Hence, the non-reciprocity might be jeopardized by a small distortion of the physical system. In order to circumvent this problem, we devised a different setup. Specifically, we considered a monolayer graphene sample in which only one half is covered by a second layer, as shown in the scheme in Fig. \ref{fig3}(b). Thus, the upper layer can only be effective in hybridizing the edge states on one side of the lower layer. In contrast, the states on the opposite edge remain quite decoupled from this `environment'. This improved setup takes advantage of both the robustness of the topological edge states of an irradiated graphene monolayer,\cite{Kitagawa2011,Perez-Piskunow2014} and the non-recyprocal Floquet band structure of the laser-illuminated bilayer system. 
As we will see it also allows for transport to be resilient to edge roughness, stacking order and disorder. 

The transmissions for the setup of Fig. \ref{fig3}(b-top) with edge roughness and $1\permil$ vacancies are shown in figure \ref{fig3}(b-bottom). It can be noticed that the directional asymmetry is preserved in this setup despite the presence of disorder and now extends over the full gap of one monolayer. Omitting edge roughness in the same sample gives results which differ very little from those in Fig. \ref{fig3}(a-bottom). Thus, directionality not only withstands imperfections but can even improve when they are added. This implies that the mechanism is \textit{anti-fragile}.~\cite{Taleb2012} Therefore, the setup of Fig. \ref{fig3}(b-top) provides a proof of concept for an electronic \textit{Floquet isolator}.

\section{Summary and final remarks.}
\textit{Isolators}, devices where transmission occurs in one direction and is suppressed on the opposite one, have been missing from the toolkit of available devices in electronics. Here we present a demonstration of this effect for the case of a laser-illuminated graphene bilayer. Laser illumination allows to introduce an effective Haldane-like term~\cite{Haldane1988} which together with the inversion symmetry breaking, produced by an electric field perpendicular to the graphene bilayer, allows for a non-reciprocal bandstructure which is exploited to produce an isolator effect.

Our starting point in this paper are the Floquet topological states produced by laser illumination on graphene. By exploiting the sublattice polarization of these edge states in zigzag-terminated samples, together with its stacking order (Bernal), we achieve a selective `switch-off' of those at one of the edges. Although this configuration is fragile against small changes in the device geometry (termination, stacking order, etc.), when covering a single edge of the system the results become much more promising: the directionality now improves with edge roughness and disorder, it is \textit{antifragile}~\cite{Taleb2012}. This implies that, rather than being merely resilient to disorder, the effect actually improves or gets better with it.

By offering a proof of concept, our work thus paves the way for new optoelectronic devices exploiting the one-wayness of the topological states for more efficient transport of energy, charge or spin. We hope that these results could estimulate the search of new ways of tailoring and harnessing non-reciprocity in electronic systems both theoretically and experimentally.

\vspace{1cm}
\textit{Acknowlegdments.} VDL acknowledges funding from CONICET. LEFFT acknowledges funding by Program ‘Inserción Académica’ 2016 of the University of Chile and FondeCyT Regular number 1170917 (Chile).

\vspace{0.2cm}
\textit{Author contributions.} LEFFT, VDL and ESM developed the concept. LEFFT and VDL designed the study. VDL wrote a code and obtained the numerical results shown here, LEFFT verified the main results. VDL and LEFFT wrote the manuscript and prepared the figures. The text was discussed and agreed by all the authors.% which was improved with comments and inputs from all authors.

\noindent
%\textbf{Contact information}\\
%\bibliographystyle{naturemag}
%\bibliographystyle{apsrev4-1} % this works
%\bibliographystyle{apsrev4-1_title}
%%\bibliographystyle{prsty}
%%\bibliographystyle{naturemag}
%\bibliography{bib}


\end{document}